# A Monte-Carlo-based and GPU-accelerated 4D-dose calculator for a pencil-beam scanning proton therapy system

Mark D. Pepin,[a] Erik Tryggestad, Hok Seum Wan Chan Tseung, Jedediah E. Johnson, Michael G. Herman, and Chris Beltran

*Department of Radiation Oncology, Mayo Clinic, 200 1st Street Southwest Rochester, MN, 55905, USA*

**Purpose:** The presence of respiratory motion during radiation treatment leads to degradation of the expected dose distribution, both for target coverage and healthy-tissue sparing, particularly for techniques like pencil-beam scanning proton therapy which have dynamic delivery systems. While tools exist to estimate this degraded four-dimensional (4D) dose, they typically have one or more deficiencies such as not including the particular effects from a dynamic delivery, using analytical dose calculations, and/or using non-physical dose-accumulation methods. This work presents a clinically-useful 4D-dose calculator that addresses each of these shortcomings.

**Methods:** To quickly compute the 4D-dose, the three main tasks of the calculator were run on graphics processing units (GPUs). These tasks were: (1) simulating the delivery of the plan using measured delivery parameters to distribute the plan amongst 4DCT phases characterizing the patient breathing, (2) using an in-house Monte Carlo simulation (MC) dose calculator to determine the dose delivered to each breathing phase, and (3) accumulating the doses from the various breathing phases onto a single phase for evaluation. The accumulation was performed by individually transferring the energy and mass of dose-grid subvoxels, a technique models the transfer of dose in a more physically realistic manner. The calculator was run on three test cases, with lung, esophagus, and liver targets, respectfully, to assess the various uncertainties in the beam-delivery simulation as well as to characterize the dose-accumulation technique.

**Results:** 4D doses were successfully computed for the three test cases with computation times ranging from 4–6 min on a server with eight NVIDIA Titan X graphics cards; the most time consuming component was the MC dose engine. The subvoxel-based dose-accumulation technique produced stable 4D-dose distributions at subvoxel scales of 0.5–1.0 mm without impairing the total computation time. The uncertainties in the beam-delivery simulation led to moderate variations of the dose-volume histograms for these cases; the variations were reduced by implementing repainting or phase-gating motion-mitigation techniques in the calculator.

---

[a] Author to whom correspondence should be addressed. Electronic mail: pepin.mark@mayo.edu.

**Conclusions:** A MC-based and GPU-accelerated 4D-dose calculator was developed to estimate the effects of respiratory motion on pencil-beam scanning proton therapy treatments. The calculator can currently be used to assess existing treatment plans and its quick runtime makes it easily usable in a future 4D-robust optimization system.

## 1. INTRODUCTION

Organ motion complicates the planning, delivery, and evaluation of external-beam radiotherapy.[1,2] This intrafractional motion, chiefly from respiration, particularly affects the treatment of tumors located in the thorax and abdomen, where organ motion up to 20 mm is possible.[3] Though non-negligible in photon treatment,[4] the dosimetric consequences of target motion are enhanced in proton therapy due to the localized energy deposition inherent to heavy-charged-particle interactions (*i.e.*, the Bragg peak).[5,6] The result is to degrade the desired treatment outcome, both in terms of target coverage and healthy-organ sparing.

Intensity-modulated proton therapy[7] delivers protons by way of pencil-beam scanning (PBS); the treatment plan consists of a collection of "spots," each containing a determined number of protons at a given particle energy, that are delivered sequentially to specified coordinates in the patient. The sequential nature of the PBS spot delivery introduces another source of motion which can interfere, both constructively and destructively, with the breathing motion in the so-called "interplay effect."[8–12] The interplay effect further distorts the delivered dose distribution, leading to a decrease in dose homogeneity that worsens with increased internal motion.[8] The intricacies of the interplay effect also mean that simply defining larger margins around the target volume is insufficient to regain target coverage as can be done in static-beam-delivery scenarios.[13]

Several methods have been developed either to reduce respiratory motion or to mitigate its effects.[14] Breath hold is a technique in which radiation is delivered while the patient holds his/her breath at a specific point in the respiratory cycle to precisely localize the position of the tumor.[15–17] During amplitude- or phase-gated treatment, the patient's breathing is monitored and radiation is delivered only during specified portions of the breathing cycle.[18–20] Tumor tracking is the most technically difficult method, in which the tumor position is monitored in real-time and the beam is actively steered to match the tumor location.[21–25] For PBS delivery, layer or volume repainting/rescanning involves delivering the planned spots multiple times, with each iteration only delivering a fraction of the total number of protons to a given spot. This serves to average out local hot/cold spots by spreading spot positional errors over different positions in the breathing cycle. Lastly, motion effects may be mitigated through four-dimensional (4D) optimization such that the motion is used as an input to the treatment-plan-creation process to develop a plan which is *a priori* more robust against motion.[26–29] Currently, our institution manages motion using breath holding, phase gating, and repainting with the goal of additionally using 4D optimization in the near future.

Faced with the complexities of 4D treatment, standard dose-calculation engines found in treatment planning systems (TPSs) that only consider dose to a static (3D) geometry are insufficient to fully describe the expected dose distribution for cases with tumor/organ motion; a 4D-dose calculator (4DDC) is thus necessary. Although some TPSs offer methods to compute 4D dose, these offerings are non-ideal for several reasons. First, TPSs generally employ analytical dose calculations instead of using the "gold-standard" of Monte Carlo simulation (MC),[30] which is particularly problematic for the heterogeneous lung geometry.[31,32] The use of analytical calculations to greatly reduce the computation time compared to CPU-based MCs is no longer needed due to the development of fast MC packages which run on graphics processing units (GPUs).[33–36] Secondly, in order to observe the effects of interplay in the final dose distribution, the dynamics of the beam's delivery must be included as a plan may not be uniformly distributed over the breathing cycle. These dynamics are facility dependent, and consequently they are not typically included in TPS 4D calculations. Lastly, studies on the mechanics of deformable dose accumulation indicate that, due to tissue deformation during the breathing cycle, one should individually accumulate energy and mass between portions of the breathing cycle as opposed to directly accumulating dose, as is typically performed in TPSs.[37–40] The purpose of this work was to develop a fast general-purpose 4DDC that accounts for beam-delivery dynamics, uses MC as the dose engine, and rigorously accumulates dose based on energy and mass. This calculator could then be used to evaluate clinical plans, to inform the effectiveness of motion-mitigation strategies, and to calculate dose for use in 4D optimization.

## 2. METHODS
## 2.A. General overview

Temporal information about patients was obtained through 4DCT imaging phase-sorted[41] into $N = 10$ 3D CT image sets ($CT_0, CT_{10}, ..., CT_{90}$ for the 0%, 10%, ..., 90% phases of the breathing cycle, respectively) by the scanner software (Siemens Healthineers, Erlangen, Germany). Note that the 4DDC was developed to be flexible in the number of phases allowed and, as such, could be used with 4DCT's sorted to a larger number of phases, ultimately limited by the available GPU memory, in future studies. The institutional standard for motion cases was to create a single 3D image set $CT_{\text{Avg}}$ by averaging all $N_g \leq N$ phases which are to be used in the treatment. $N_g = N$ for free-breathing deliveries while $N_g < N$ for gated deliveries where the portion of the breathing cycle with the $N_g$ phases is termed the "gate-on window." Structure contouring and treatment planning were then performed on $CT_{\text{Avg}}$. A full structure set thus existed for $CT_{\text{Avg}}$, but usually not for the individual breathing phases. However, a minimal structure set containing at least an "external" contour is required by the MC for each breathing phase (to limit the applicable bounds of particle transport and computation time). The external contour from $CT_{\text{Avg}}$ was rigidly transferred to the individual breathing phases using the MIM software package (MIM Software Inc., Cleveland, OH). The nominal external contour consisted of the entire patient body, treatment table, and immobilization device plus an additional margin.

Dose accumulation generally requires deformation vector fields (DVFs) determined by deformable image registration (DIR). The DVFs map points on the various breathing phases to their corresponding locations, accounting for deformation, on a designated accumulation phase. The accumulation phase was a variable input to the calculator; for the studies considered in this work, the more-stable near-end-of-exhalation phase[42] $CT_{50}$ was adopted. DIR was performed using MIM, which uses the VoxAlign Deformation Engine™ to perform intensity-based freeform deformable registrations.[43] This resulted in $N_g - 1$ DVFs which were imported to the 4DDC as DICOM registration files. Note that, since the 4DDC was developed to accept and parse DICOM input, DIR could have been performed with any DICOM-compliant registration system.

The DICOM input to the 4DDC is then summarized as: $N_g$ 4DCT-based CT image series, $N_g$ structure set files, $N_g - 1$ registration files, and one treatment-plan file to specify the treatment fields and spot characteristics (location, energy, weight). These DICOM files were processed using a PYTHON script to filter inputs for the 4DDC. After the DICOM files were properly parsed, the 4DDC preformed three GPU-based computations, which are discussed in detail in the following subsections. These steps were simulating the beam delivery (Sec. 2.B), calculating the dose via MC (Sec. 2.C), and accumulating dose across the breathing phases onto $CT_{50}$ (Sec. 2.D). To be DICOM compliant, and clinically usable, the final product of the 4DDC should be a DICOM dose file which can be imported and viewed in any TPS. As such, the last step of the 4DDC was another PYTHON script which converted the final accumulated-dose distribution on $CT_{50}$ to a DICOM dose file.

## 2.B. Beam-delivery simulation

The beam-delivery simulation determined what portions of the treatment plan would be delivered during each breathing phase. This was accomplished by simultaneously tracking the timing of the patient breathing, to indicate the current breathing phase, and the timing of the plan delivery. The end result of these simulations was a set of $N_g$ subplans, per field, indicating the number and weighting of spots which would be delivered on each breathing phase. The beam-delivery simulation was performed in a parallel manner on a GPU, using a CUDA C kernel, with all fields simultaneously simulated.

The 4DDC was designed to be a flexible tool by giving the user control over many aspects of the delivery simulation mimicking variables and options found in the clinic. The user could specify whether to perform a phase-gated delivery and define the associated beam-on window. The plan could also be reorganized to create an isolayered-repainted[44] delivery with a specified maximum monitor unit (MU) per spot delivery. A source of uncertainty in the delivery was the breathing phase during which delivery commences, which the user could either specify or allow to be randomly selected.

The length of each breathing phase $t_{ph}$ was characterized in the simulation by a user-input breathing period mean $\mu$ and standard deviation $\sigma$. If the breathing were perfectly uniform, $t_{ph}$ would be $t_{ph} = \mu/N$. As patient breathing patterns are not perfectly uniform, $t_{ph}$ was instead a random variable described as $t_{ph} \sim \mathcal{N}(\mu,\sigma)/N$, *i.e.*, it was found by sampling a normal distribution $\mathcal{N}(\mu,\sigma)$, characterized by $\mu$ and $\sigma$, and then dividing by the total number of phases.[b] This sampling allowed for some measure of irregular breathing, without requiring an imported breathing trace or 4D-MRI. Clinically, the appropriate values for $\mu$ and $\sigma$ for a given patient could be obtained by measuring a breathing trace or from the 4DCT reconstruction software. For the studies presented in this manuscript, these values were treated as input variables whose values affect the overall uncertainty in the 4DDC result.

Our institution employs a Hitachi ProbeatV5 synchrotron-based system with 97 discrete energy levels ranging from 71.3 to 228.8 MeV. The timing for spot delivery from such a system was well characterized in Ref. 45 and is summarized in App. A. On the assumption that the characterized system behaves similarly to ours, the values given in Ref. 45 were used for the 4DDC timing simulation with one exception. When beam extraction begins, the initial 70 ms of the beam are discarded as it is of poor quality; when the delivery is non-gated, this occurs immediately after resetting the synchrotron. In a gated delivery, however, after the synchrotron resets it can be held in the accelerator until receiving a gate-on signal, at which point the 70 ms of beam is discarded. Only non-gated deliveries were considered in Ref. 45 and, as such, this time was combined into the synchrotron-reset time. For the 4DDC, the beam-discarding time was separately tracked to allow for gated deliveries. The 4DDC also optionally allowed for uncertainty in these delivery parameters; the uncertainties were either provided in Ref. 45 or obtained from the authors themselves. Uncertainty values were sampled from a Gaussian distribution with mean and standard deviation corresponding to the central value and uncertainty for the given parameter.

---

[b] All random-number sampling on the GPU was performed using the cuRAND library.

The outcome of the beam-delivery simulation, and thus the final 4D-dose distribution and dose-volume histograms (DVHs), could be affected by the three sources of uncertainty identified above. The effect of these uncertainties was considered by running 25 trials of the 4DDC where, for each trial, the initial-breathing phase, breathing period, and synchrotron delivery parameters were simultaneously varied according to their statistical distributions. These distributions were: starting phase $\sim \mathcal{U}(CT_0, CT_{90})$, synchrotron parameters $\sim \mathcal{N}(\mu, \sigma)$ (as described above), and breathing period $\sim \mathcal{U}(3,7)$ s. Here, $\sim \mathcal{U}(a,b)$ indicates random selection from a discrete uniform distribution inclusively spanning from $a$ to $b$. These 25 trials were also used to characterize the timing of the GPU-based components of the 4DDC. The effect of motion mitigation was considered by repeating the 25 trials using a gated delivery with protons delivered on $CT_{30}$–$CT_{70}$ (50% duty cycle) and then again for free breathing with a max MU of 0.005 MU for isolayered repainting.

## 2.C. Dose calculation MC

The GPU-based MC dose-calculation engine used by the 4DDC has been previously described in Refs. 33 and 34. The simulation is a class II condensed-history proton transport package which is capable of simulating $1 \times 10^7$ protons in less than 30 s on a gaming graphics card. As discussed in Sec. 2.D, the dose-accumulation algorithm requires knowledge of the tissue density of each CT voxel. This information is already calculated internally by the MC, converting Hounsfield Units to material density,[46] and minimal changes to the MC package were required to extract this information as an additional output. The outputs from a single MC run were the dose for and the density of each original CT voxel which received non-zero dose during the simulation. In these simulations, the dose grid (3D bins into which dose is counted) was identical to the CT-voxel grid, which is not the case in most TPS calculations where the dose grid may have coarser resolution. For the 4DDC, $N_g$ simulations were performed per field. The 4DDC distributed the MC jobs over all available GPU devices on the system; the results presented here were run on a research system with eight NVIDIA (NVIDIA Corporation, Santa Clara, CA) GeForce GTX Titan X devices.

## 2.D. Dose accumulation
### 2.D.1. DVF calculation

The end result of any DICOM-compliant DIR software is a deformable spatial registration DICOM file which encodes a DVF as detailed in App. B. The DVF transformation can be mathematically expressed as

$$\boldsymbol{r}_f = T^{0 \to f}(\boldsymbol{r}_0), \tag{1}$$

where $\boldsymbol{r}_0$ is a location on the initial image $I_0$, $\boldsymbol{r}_f$ is the mapped location on the final image $I_f$, and $T^{0 \to f}(\boldsymbol{r}_0)$ is a function which computes the DVF translation at $\boldsymbol{r}_0$. In the DICOM file, the DVF is encoded as a three-dimensional grid on $I_0$ along with the translations at the centers of this grid's voxels. Note that, to reduce file size, only minimal information about this grid is actually stored in the file (see App. B). For the 4D-breathing problem, a DVF was needed for each non-accumulation phase in the gating window. Note that whether an accumulation phase corresponds to $I_0$ or $I_f$ in Eq. (1) depends on the deformable-dose summation technique used (see Sec. 2.D.2); instead, the terminology of Ref. 40 is adopted here such that the accumulation phase is called the reference image $I_R$ and all other phases from which dose is accumulated are called source images $I_S$'s.

To use the DVF information for dose accumulation, the 4DDC needed first to calculate the full DVF

based on the stored minimal-grid information and then to interpolate between deformation-grid voxel locations to derive the deformation at dose-grid (CT voxels) locations. These steps were performed in two separate CUDA C calls on the GPU, the first of which served to construct the DVFs for all $I_S$'s; running on a one-thread-per-deformation-grid-voxel basis to calculate $r_f$ for all voxels. When this kernel finished, the computed values for $r_f$ were reformatted and bound to 3D texture memory on the GPU. Texture memory is a special type of device memory allowed in CUDA C which increases performance for applications with spatial locality (nearby threads accessing addresses which are themselves near each other) and which provides linear interpolation (trilinear for 3D textures). Because nearby CT voxels will typically have similar deformations, both of these properties made textures desirable for use here. The second CUDA C kernel accumulated the MC-calculated doses from each breathing phase onto $CT_{50}$. However, a discussion of dose accumulation techniques is first required before detailing this GPU kernel.

## *2.D.2. Deformable dose accumulation*

There is a large body of literature surrounding various techniques for performing deformable dose summation.[37–40,47–54] In all methodologies, the final dose $D_i$ for voxel $i$ on $I_R$ is

$$D_i = D_i^R + \sum_S D_i^S, \qquad (2)$$

where $D_i^S$ is the dose from source phase $S$ which maps to voxel $i$ and $D_i^R$ is the dose originally delivered to $I_R$. The difference between methodologies lies in the determination of $D_i^S$. Note that, for the 4DDC, any relative weighting between the phases was accounted for by the beam-delivery simulation, and thus there were no additional weight factors between the phases in Eq. (2). The various methods can be sorted into three general categories:[47] dose-interpolation methods (DIMs), energy-/mass-transfer methods (EMTMs), and voxel-warping methods (VWMs).[48,49,55] Both a DIM and an EMTM were implemented in the 4DDC.

The general approach for DIMs considers each voxel on $I_R$ by:[47,51,52,54,56] determining where the voxel's location maps to on each $I_S$, finding the appropriate doses at these locations, pulling these doses back to $I_R$, and assigning them to the given voxel. This approach guarantees that every voxel on $I_R$ receives a mapped dose, but, as there is generally no one-to-one correspondence between voxels on $I_R$ and any given $I_S$,[38,40] it does not guarantee that all dose-filled voxels on the $I_S$'s are pulled back or that the rearrangement of mass is properly handled. The method implemented in the 4DDC is the trilinear-interpolation method of Ref. 54 and is illustrated in Fig. 1. For each source phase $S$: each voxel on $I_R$ was split into $n \times n \times n$ subvoxels, the central location of each subvoxel $r_{i_k}$ was mapped to $I_S$ giving $T^{R \to S}(r_{i_k})$, the trilinearly-interpolated dose at this location $\widetilde{D}^S[T^{R \to S}(r_{i_k})]$ was computed and the final $D_i^S$ was then

$$D_i^S = \frac{1}{n^3} \sum_{k=1}^{n^3} \widetilde{D}^S[T^{R \to S}(r_{i_k})]. \tag{3}$$

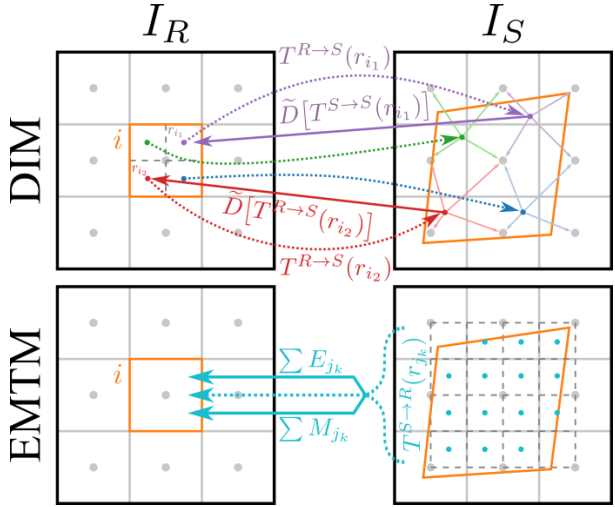

Fig. 1 Two-dimensional representation of the dose-accumulation methods implemented in the 4DDC showing the DIM (top) and EMTM (bottom) techniques. In each case, dose is accumulated onto the selected voxel $i$ (highlighted in solid orange) on $I_R$ (left) from a single source phase $I_S$ (right) where the voxel warps to encompass a larger area. The DIM technique divides $i$ into subvoxels (dashed lines) and maps (dotted lines) their centers $r_{i_k}$ (small dots) to $I_S$. The dose at the mapped location is calculated by interpolating nearby dose-grid locations on $I_S$ and pulled back (solid lines, only two shown for clarity) to $I_R$. The total dose in $i$ is then the average of all interpolated subvoxel doses. The EMTM technique instead subdivides the $j$ voxels on $I_S$ and considers all subvoxels whose centers $r_{j_k}$ map to $i$. The energy and mass from all such subvoxels are individually summed with the final dose in $i$ being the ratio of the summed energy to the summed mass.

Even with the subvoxelization and interpolation above, fundamental issues with pulling dose based on $I_R$'s voxels remain (as demonstrated in Refs. 38 and 40) due to voxels merging and/or splitting between the images. Instead, EMTMs aim to explicitly use the definition of dose by decoupling the tracking of energy and mass during the deformation and pushing them from the $I_S$'s to $I_R$.[37–40] A defining characteristic of EMTMs is that, to push the energy/mass, the DVFs used go in the opposite direction as those used in DIMs. The method implemented in the 4DDC was similar to that of Ref. 37 and is also illustrated in Fig. 1. As in the DIM implementation, the initial voxels were split into $n^3$ subvoxels but, in this case, the initial voxels were from each $I_S$. The mass of and the energy deposited in the $k^{\text{th}}$ subvoxel of voxel $j$ ($M_{j_k}$ and $E_{j_k}$, respectively) were first calculated as

$$M_{j_k} = \rho_j v_k, \qquad E_{j_k} = M_{j_k} D_j, \tag{4}$$

where $D_j$ is the dose in voxel $j$, $\rho_j$ is its density as calculated by the MC (see Sec. 2.C), and $v_k$ is the volume of the subvoxel (same for all subvoxels), and. The dose to voxel $i$ from phase $S$ was then computed as

$$D_i^S = \frac{\sum_{j_k \in \mathbb{J}} E_{j_k}}{\sum_{j_k \in \mathbb{J}} M_{j_k}}, \tag{5}$$

with the mathematical set $\mathbb{J}$ defined as

$$\mathbb{J} \equiv \{j_k | T^{S \to R}(\boldsymbol{r}_{j_k}) \in V_i\}, \tag{6}$$

that is, the set of all subvoxels whose center $\boldsymbol{r}_{j_k}$ maps into the volume $V_i$ contained by voxel $i$. The authors of Ref. 37 categorized this method with DIMs as a "dose mapping" instead of an "energy transfer" model as it uses a discretized dose at the fundamental level (they describe it as a mass-weighted dose). However, given the form of Eq. (5) and the directionality of the required DVFs, it is here categorized with the EMTMs instead of the DIMs. Note also that those authors compared this method to other accumulation techniques, including the DIM of Eq. (3), and found this style of EMTM to be superior; the EMTM technique was thus the default technique in the 4DDC.

Both the DIM from Eq. (3) and the EMTM from Eqs. (4)–(6) were implemented in the 4DDC to run on the GPU. Input options allowed the user to choose both the accumulation method and the voxel-subdivision scale $n$ which was used by either method. The GPU implementation was such that all of the dose files generated by the MC were read in and transferred to the GPU as 3D texture memories; the built-in trilinear interpolation of textures was used for the DIM technique. The computation then proceeded with a single main-voxel-per-thread approach; for the DIM, each voxel on $CT_{50}$ was assigned a thread per non-accumulation phase while, for the EMTM, each voxel on a non-accumulation phase was assigned a thread. Assigning a thread per each subvoxel for all phases would require more memory than available on the GPU, thus looping over subvoxels was performed on the device which could slow the kernel for large $n$. To avoid race conditions in the EMTM parallelization, CUDA C atomic functions were used when summing energy and mass. After the initial kernel for either method was run, a second kernel was called, with one thread per voxel on $I_R$, to perform the final dose summation of Eq. (2). The final accumulated dose was then written to a binary file such that it could be converted to a DICOM dose file as described in Sec. 2.A. Since the DVFs typically cover less than the full CT geometry, a mask was applied to only report dose in the volume encompassed by all DVFs to avoid artifacts in the 4D dose.

The determination of an appropriate value for the voxel-subdivision scale $n$ is presented in the results and discussion sections. This determination considered the facts that at larger values of $n$ the 4D-dose distribution should be more stable but there is also the potential at large $n$ for the dose-accumulation component to become overly time consuming (because each GPU thread loops over all subvoxels for a given primary voxel). As such, both the convergence of DVHs and the time to complete the dose accumulation were studied as functions of $n$ to determine an optimal scale of voxel subdivision.

### 2.E. 3D comparison tests

Two preliminary tests were performed to benchmark the 4DDC against known 3D situations. The first involved creating a mock 4D data set by using the same 3D CT image set for all breathing phases coupled with a manually-entered "identity" DVF which mapped each point to itself. Since each breathing phase now had an identical geometry, and there was no warping applied between phases, the result should match the 3D-static dose as found by running the original MC calculator on $CT_{50}$. The second test was to give $CT_{50}$ as the only phase to the 4DDC. In this case, the beam-delivery simulation should place the entire plan on $CT_{50}$ as if it were a non-motion site and thus require no dose warping. The results of both tests gave the expected match with the 3D-static dose.

### 2.F. Patient test-case studies

Three retrospective test cases were used to study the results and characterize the various uses of the 4DDC. The patients were selected to cover a variety of tumor locations with the patients having lung, esophagus, and liver targets, respectively. Treatment plans were originally created (on $CT_{Avg}$) using a commercially available TPS (Eclipse, Varian Medical Systems, Palo Alto, CA) at the time of treatment. The patients were treated with no active motion mitigation (free breathing), however the plans were delivered with isolayered repainting[44] with a maximum MU per spot delivery of 0.005 MU. To assess the characteristics of the 4DDC, and the effects of repainted and gated deliveries, alternative plans were considered where the original plans were recast to the non-repainting maximum per spot delivery of 0.1 MU. Characteristics of the plans and geometries for the cases are given in Table I.

| Characteristic | Lung | Esophagus | Liver |
|---|---|---|---|
| Prescription [Gy] | 60 | 50 | 67.5 |
| Fields | 3 | 2 | 3 |
| Tumor motion [mm] | 5.4 | 5.3 | 7.5 |
| Tumor volume [cm³] | 79.6 | 204.4 | 581.0 |
| CT voxel size [mm] | $1.27 \times 1.27 \times 2.00$ | $1.27 \times 1.27 \times 2.00$ | $1.27 \times 1.27 \times 1.50$ |
| CT voxel grid size | $427 \times 246 \times 161$ | $430 \times 272 \times 235$ | $452 \times 307 \times 281$ |
| DVF voxel size [mm] | $2.55 \times 2.54 \times 2.56$ | $2.55 \times 2.55 \times 2.57$ | $2.55 \times 2.54 \times 2.57$ |
| DVF voxel grid size | $175 \times 109 \times 126$ | $184 \times 122 \times 218$ | $184 \times 135 \times 164$ |

Table I Characteristics of the plans and geometries for the test patients considered with the 4DDC. Sets of three numbers give the appropriate values in the $x \times y \times z$ dimensions, respectively. Tumor motion was estimated as the maximum displacement, between any two phases, of the centroid of the target's volume while the tumor volume was estimated as the target's mean volume across all phases. CT voxel sizes were the standard resolutions used for these treatment sites while the DVF voxel size and grid were automatically determined by MIM, the DIR software used.

For a full analysis, region-of-interest (ROI) contours were required on $CT_{50}$ for assessment of the accumulated-dose distribution under various scenarios. For each case, one healthy-tissue ROI and the clinical target volume(s) (CTVs) were identified on $CT_{50}$. The esophagus and liver cases had high- and low-dose CTVs while the lung case had a single CTV. The healthy-tissue ROIs for the cases were the esophagus, heart, and non-target liver tissue for the lung, esophagus, and liver cases, respectively. For all scenarios studied with the 4DDC, the DVH for each ROI was evaluated.

## 3. RESULTS
### 3.A. Baseline results

The 4D-accumulated dose distribution on $CT_{50}$ for the lung case is shown in Fig. 2. The parameters used for the dose shown in the figure will be called the "4D-baseline" values which correspond to using all ten phases (a non-gated delivery), a starting phase of $CT_0$ (arbitrarily selected), a breathing period of $5 \pm 0.5$ s (arbitrarily selected), the nominal experimental values for the synchrotron delivery parameters (from Ref. 45), and EMTM dose accumulation with a voxel-subdivision scale of $n = 5$ (see Sec. 4.A).

Fig. 2 also gives the portions of the 4D-accumulated dose that were delivered to $CT_{50}$ and $CT_0$ (which has been deformed to $CT_{50}$ in the figure), highlighting the variation between phases. Additionally, the 4D-baseline dose is compared to the 3D-static dose (as calculated with the second test described in Sec. 2.E). The 3D to 4D comparison serves to give a sense of the effects of motion on the distribution.

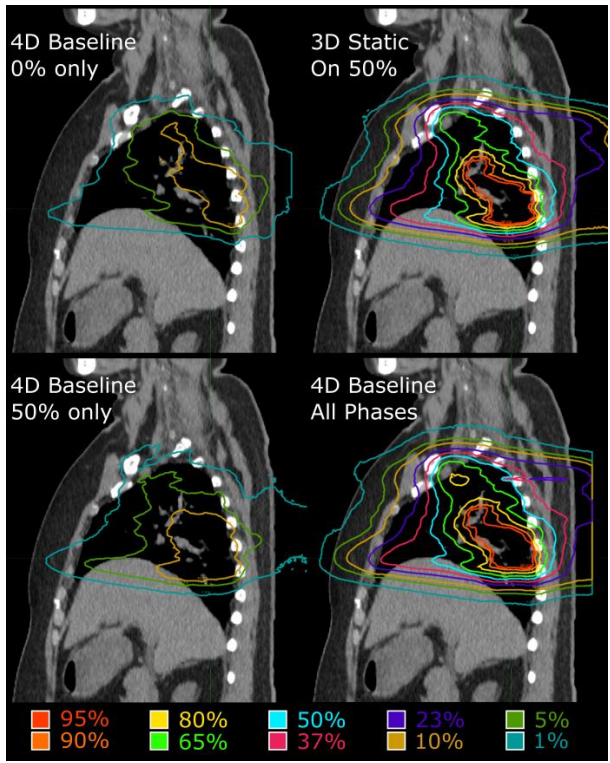

Fig. 2 Isodose lines (percentage of prescription) on $CT_{50}$ for the lung case showing the dose distribution from the 4D-baseline (all phases deformed to and summed on $CT_{50}$) and 3D-static deliveries. Also shown are the components of the 4D-baseline dose which were delivered on $CT_0$ and $CT_{50}$ individually, with the dose delivered to $CT_0$ presented after warping it to $CT_{50}$. The differences on the far-right sides of the distributions are due to the DVF-defined mask applied in the 4DDC which is not applied for the 3D-static or $CT_{50}$ only cases, which use no DVF.

### 3.B. Dose-accumulation characterization

The 4DDC was characterized as a function of voxel-subdivision scale $n$ (recall that each primary voxel was split into $n^3$ subvoxels during dose-accumulation) while otherwise using the 4D-baseline values for the delivery parameters. Fig. 3 shows DVHs for both the EMTM and DIM accumulation techniques for $n = 1, 2,$ and $100$ as a test for convergence in the 4D-dose distribution. For the DIM case, the curves are consistent regardless of the degree of subvoxelization. For the EMTM case, the DVHs with no subdivision ($n = 1$) differ substantially from those with subdivision. However, the equivalence of the $n = 2$ and $100$ curves indicates convergence already with eight subvoxels. This degree of subdivision corresponds to subvoxel dimensions of 0.5–1.0 mm as indicated by Table I.

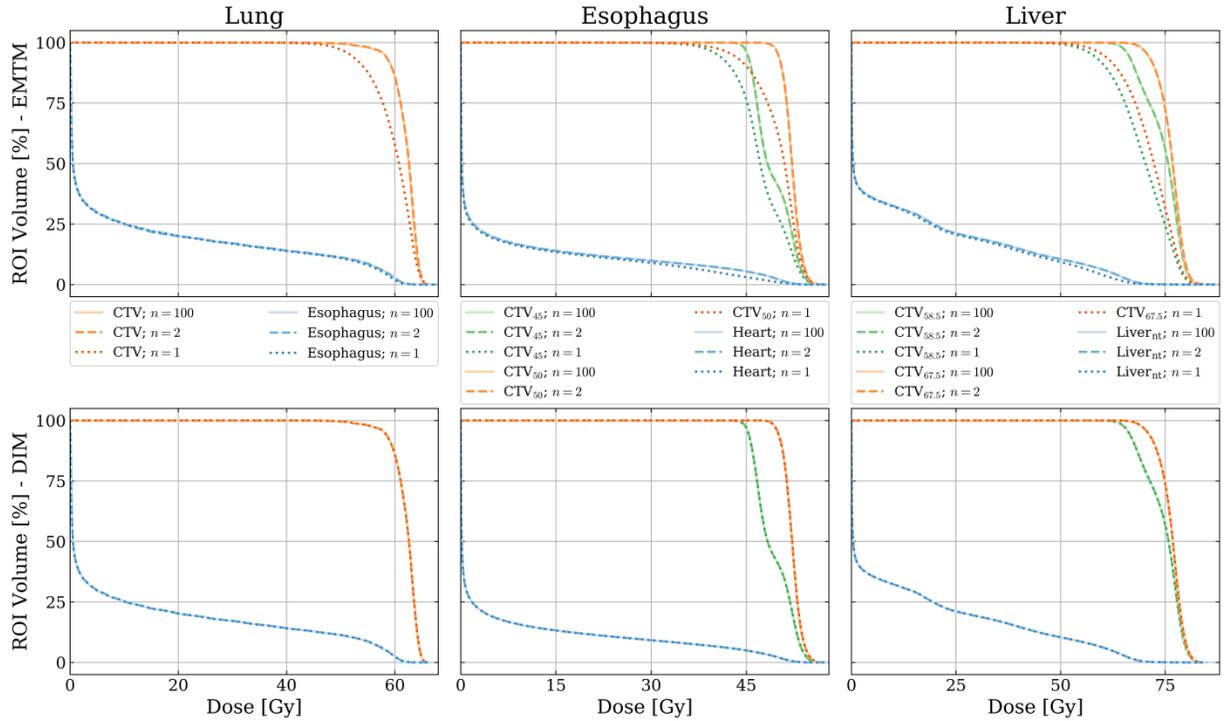

Fig. 3 DVHs for all ROIs for the lung (left), esophagus (middle), and liver (right) cases using EMTM (top) and DIM (bottom) for dose accumulation under various voxel-subdivision scales $n$. Captions in the middle apply to both top and bottom plots. For the DIM, the DVHs computed with no voxel subdivision ($n = 1$) are already consistent with those of $n = 100$. For the EMTM, the DVHs for $n = 1$ substantially differ, however, convergence with $n = 100$ is already seen by $n = 2$. Abbreviations – $CTV_X$: CTV at a prescription of $X$ Gy, $Liver_{nt}$: non-target liver.

The time to complete the accumulation step for each patient and accumulation technique is given in Fig. 4. For $n \lesssim 5$, the time is relatively flat and is dominated by file input/output. The EMTM technique requires slightly more time here as it must load the voxel densities in addition to the doses. For $n \gtrsim 5$, the time varies as $n^3$ which is indicative of looping over subvoxels. The EMTM technique is quicker in this regime due to internally controlled differences in how the GPU caches texture memory when sampling the DVF.

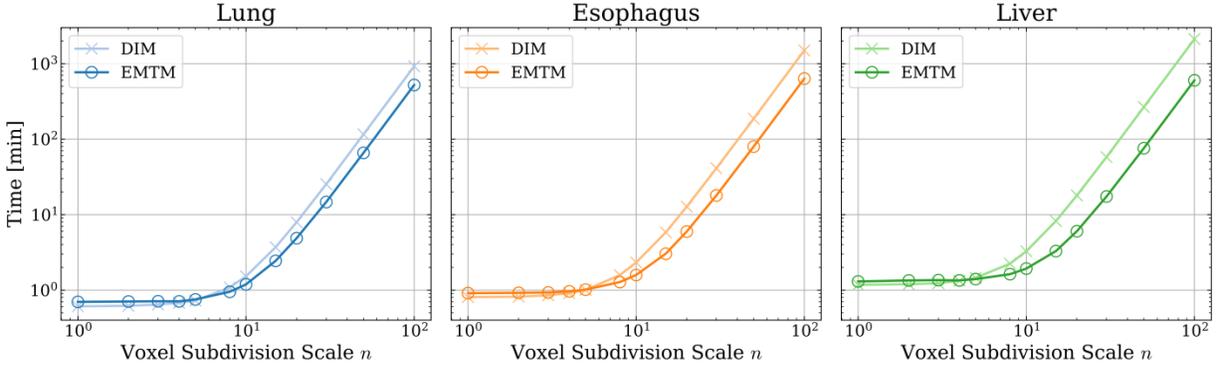

Fig. 4 Time to complete the dose-accumulation stage of the 4DDC as a function of voxel-subdivision scale $n$. The flat behavior at low $n$ is the file input/output time; the EMTM additionally reads the density file. The $n^3$ behavior at high $n$ is due to looping over subvoxels on the GPU threads.

### 3.C. Delivery simulation characterization

The effects on the DVH curves from sampling the beam-delivery uncertainties 25 times are seen in Fig. 5. Mild variations in the target DVHs are seen with the breathing period having the largest effect, followed by the starting phase selection. The times to complete the GPU-based components of the 4DDC based on these 25 trials are given in Table II. The largest contributor to the time was the particle MC at 2.5–3.0 min, followed by the dose-accumulation step at 1–2 min while the beam-delivery simulation took only ∼7 s. The effects of repeating the sampling of beam-delivery uncertainties while instead using a gated or repainted delivery are shown in Fig. 6. The 0.005 MU repainted value is the value used clinically (see Sec. 2.F) and thus the repainted trials represent the original treatment plan.

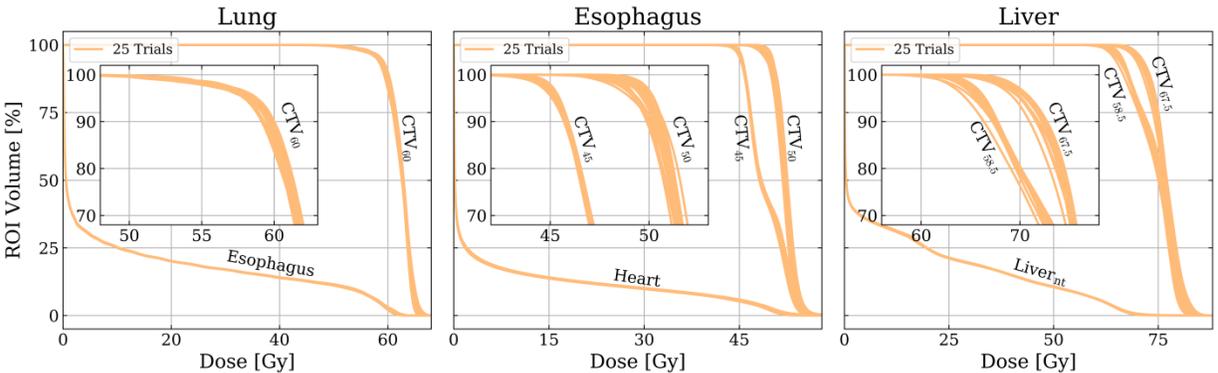

Fig. 5 DVHs for all ROIs for the lung (left), esophagus (middle), and liver (right) cases for 25 different uncertainty trials. For each trial, the starting phase, breathing period, and synchrotron delivery parameters were sampled according to their uncertainties (see main text). The text labels indicate the grouping of curves corresponding to the given ROI; the width of a particular grouping is indicative of the uncertainty for that ROI. Abbreviations – $CTV_X$: CTV at a prescription of $X$ Gy, $Liver_{nt}$: non-target liver.

| Component | Lung [s] | Esophagus [s] | Liver [s] |
|---|---|---|---|
| Beam-Delivery Sim. | $6.7 \pm 0.1$ | $7.0 \pm 0.1$ | $7.4 \pm 0.1$ |
| Particle MC | $156.8 \pm 2.1$ | $135.0 \pm 4.4$ | $174.9 \pm 2.1$ |
| Dose Accumulation | $45.2 \pm 0.7$ | $90.2 \pm 0.5$ | $121.9 \pm 1.7$ |
| Total | $245.8 \pm 2.2$ | $297.4 \pm 12.2$ | $367.3 \pm 9.3$ |

Table II Time, in seconds, to complete the main components of the 4DDC as well as the total time for each of the three cases. The beam-delivery simulation and dose accumulation were performed on a single GPU while the particle MC was performed across eight GPUs. Time not accounted for in the given components is due to DICOM-file input/output and transferring data between GPU components.

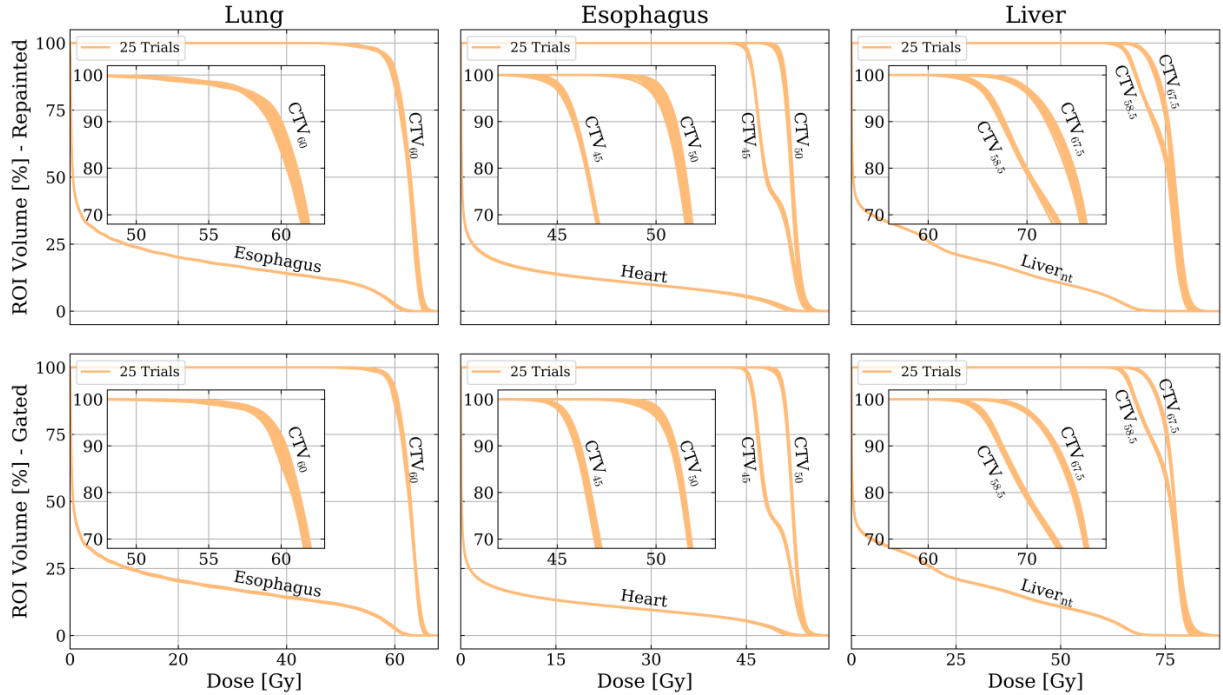

Fig. 6 DVHs for all ROIs for the lung (left), esophagus (middle), and liver (right) cases for 25 different uncertainty trials each for isolayered-repainted (top) and gated (bottom) deliveries. The text labels indicate the grouping of curves corresponding to the given ROI; the width of a particular grouping is indicative of the uncertainty for that ROI. The uncertainties are reduced relative to Fig. 5. Abbreviations – $\text{CTV}_X$: CTV at a prescription of $X$ Gy, $\text{Liver}_{nt}$: non-target liver.

## 4. DISCUSSION

### 4.A. Characterization and performance

Selecting an optimal value for the voxel-subdivision scale was done by comparing Fig. 3 and Fig. 4; although dose convergence was seen at $n = 2$, there was no time penalty until $n \simeq 5$. Because further subdivision should, in principle, provide more accurate results, $n = 5$ was selected as the optimal value and used for the 4D baseline as described in Sec. 3.A. The uncertainties in the beam-delivery simulation created mild variations in the target DVHs and, as expected, using a motion mitigation technique reduced the incidence of lower-coverage curves and tightened the overall distributions.

Using the optimal value of $n$, the most time consuming component of the 4DDC was the particle MC as seen in Table II. However, as discussed in Sec. 2.C, the MC was distributed amongst eight GPUs and thus this time could be reduced by spreading the MC jobs amongst more GPUs. The differences in timing between patients are related to the voxel-grid sizes and number of fields as given in Table I; larger treatment areas with sparser dose distributions lead to more voxels with non-zero dose and less efficient texture caching, while more fields lead to more MC runs and subsequently more dose files to be loaded by the accumulation component.

## 4.B. Uncertainties and limitations

The 4DDC can only be as accurate as the assumptions that its underlying components are based on. For the beam-delivery simulation, this includes the characterization of the synchrotron parameters. The values from Ref. 45 used in the simulation were not measured with our institution's synchrotron and, although that system is the same Hitachi design and configuration, no two systems are entirely identical and it is an assumption that these parameters are applicable to the our synchrotron. Additionally, variation in the beam current is seen in the day-to-day operations of the synchrotron and although some uncertainty on the current was studied above, it cannot fully encapsulate the observed stochastic variation. Lastly, the delivery simulation was developed assuming that a full synchrotron reset is performed between all energy layers, so-called "single-energy extraction." However, an alternative mode where protons can be decelerated within the synchrotron such that a full reset is not required when stepping down in energy, so-called "multi-energy extraction," is currently being used. Delivering such a plan would require moderate modifications to the existing beam-delivery simulation.

There are also inherent limitations with regard to 4DCTs, DIRs, and EMTM-style dose accumulation. This study phase-sorted the 4DCTs into 10 phases, however the type of sorting[41] and number of phases used[57] have both been shown to affect the resulting 4D-dose distribution; some studies indicate that using more phases and/or interpolating between phases provides more accurate 4D-dose assessment.[58] Many different DIR algorithms exist with variations between metrics and transformation approaches,[59] and these differences can have dosimetric effects: the CTV volume receiving 95% of the maximum dose can vary by 5–10% between DIR algorithms.[60] For dose accumulation, DIMs and EMTMs use DVFs with opposite directionality and thus a true comparison between these methods should use a DIR method which is analytically invertible. The free-form style transformation of MIM used for this study is not invertible, but, as a systematic comparison is beyond the scope of this work (see, *e.g.*, Ref. 37) this small effect was not considered. The EMTM-style dose accumulation presented above explicitly uses the density of each voxel, and any uncertainty in its calculation from Hounsfield units would necessarily impact the dose-accumulation results.

### 4.C. Current and future uses

Given the results in Sec. 3.C and the discussion above, it needs to be emphasized that a single run of the 4DDC should not be interpreted as giving the exact 4D dose. Instead, it should be viewed as giving an estimate of the scale of the effect of motion and multiple runs of the 4DDC should be used to gauge the uncertainty in that estimate. Notwithstanding the above, the results of the 4DDC are still clinically relevant as they give a more accurate assessment of the effects from motion, particularly those from interplay effects, than simple solutions commonly used such as calculating the dose statically on only $CT_0$ and/or $CT_{50}$. Experimental validation of any 4DDC will increase confidence in its use (see, *e.g.*, Ref. 61) and studies for such validation of the presented 4DDC are underway.

As indicated in Sec. 1, 4D optimization is a more advanced technique for mitigating the effects of motion. In particular, an in-house MC-based and GPU-accelerated robust physical- and biological-dose optimizer has been developed to quickly generate treatment plans.[62,63] Given the shared MC and GPU framework between this optimizer and the presented 4DDC, integrating the two together into a 4D-robust optimizer is a future direction for study.

### 5. CONCLUSIONS

This work presented a fast, clinically relevant, 4D-dose calculator which is based on MC-simulated dose. This is accomplished by three GPU-accelerated tasks which respectively simulate the delivery of a plan amongst 4DCT breathing phases to generate subplans on each phase, calculate the dose of each subplan via MC simulation, and accumulate the subdoses onto the near-end-exhale phase via DVFs. This calculator was characterized and tested with three retrospective test cases, where stable 4D-dose distributions were computed in 4–6 min. The next steps include experimental validation and development of a 4D-robust optimizer to generate treatment plans which are less sensitive to motion.

## ACKNOWLEDGMENTS

The authors thank J. Shen for sharing details from his studies with synchrotron-delivery timing. This work was supported by a grant from Varian Medical Systems.

## CONFLICTS OF INTEREST

The authors have no relevant conflicts of interest to disclose.

## A. DESCRIPTION OF SYCHROTRON DELIVERY OF PROTONS

As mentioned in Sec. 1, in PBS the beam is delivered as individual spots which each have a specific $(x, y)$ position and energy; all spots of the same energy form a "layer." A general schematic of the relevant timing parameters to deliver a layer from a synchrotron is shown in Fig. 7 and proceeds as follows: low energy protons are injected into the synchrotron and then accelerated to the desired energy. Once at energy, there is a short preparation period, to check the beam energy and prepare the steering magnets, before either holding the beam in the accelerator to wait for a gate-on signal or immediately proceeding to beam extraction. When extraction does begin, a second preparation period occurs to discard the initial 70 ms of low-quality beam. Extraction ceases when one of a number of conditions is met: all spots of the given layer are completed, the charge capacity of the synchrotron (∼2 nC) is depleted, or the flat-top and/or extraction time limits (∼8 s) are exceeded. The last two conditions are imposed as the phase-space quality of the remaining protons in the synchrotron can only be maintained for so long. Once extraction ends, preparation is made for deceleration followed by deceleration itself and then the beginning of a new cycle. The time to reset the synchrotron between layers, including the preparation phases which precede and follow the reset, is energy dependent and grouped together in the "layer-switch time" (∼2 s).

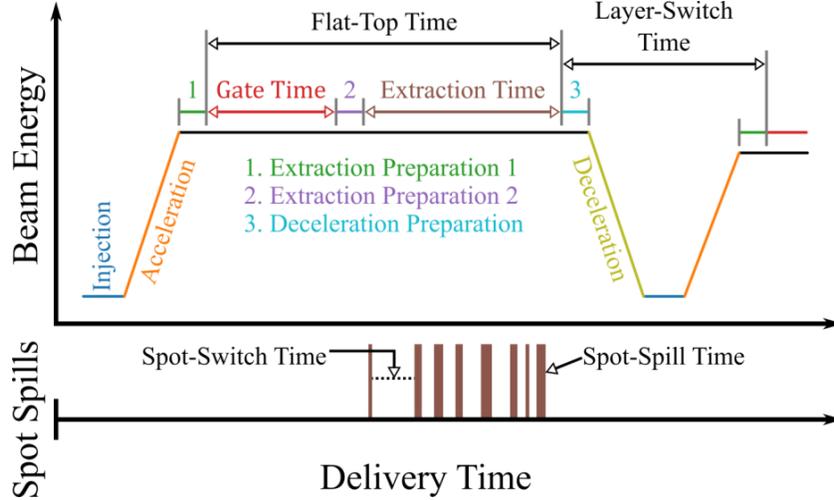

Fig. 7. Schematic of the synchrotron delivery pattern demonstrating the various components of the delivery timing for (top) the beam energy profile and (bottom) the delivery of individual spots. Low-energy protons are injected into the synchrotron, accelerated to the desired energy, held there until the energy layer is complete or a parameter limit is exceed, and then decelerated. While at energy, time is spent undergoing three short preparation periods, waiting for a gate signal, and extracting protons. During extraction, spots are delivered as individual spills which require a finite time to deliver (width of vertical bars) and switch between. Various quantities relevant to the 4DDC computation are also indicated, including the flat-top, extraction, gate-holding, and layer-reset times.

While at a given energy, there is a certain beam current (MU/s) which dictates the rate at which protons are spilled from the synchrotron and, therefore, there is a finite time, called the "spot-spill time," required to deliver a spot of any given MU at any given energy. It also takes finite time, called the "spot-switch time" to redirect the beam between spots (during which the beam is off) which is dictated by the speed of steering magnets and the time to prepare and verify the magnet operation before and after repositioning the beam. See Ref. 45 for further details and specific values for each timing component.

## B. DEFORMATION VECTOR FIELD CALCULATION

The DICOM standard for encoding DVFs, as would be computed by any compliant DIR software, is given by[64]

$$\begin{bmatrix} x_f \\ y_f \\ z_f \\ 1 \end{bmatrix} = M_{\text{Post}} \left( M_{\text{Pre}} \begin{bmatrix} x_0 \\ y_0 \\ z_0 \\ 1 \end{bmatrix} + \begin{bmatrix} \Delta x \\ \Delta y \\ \Delta z \\ 0 \end{bmatrix} \right), \qquad (7)$$

where $(x_0, y_0, z_0)$ is the location of the center of a DVF-grid voxel on the initial image $I_0$, $(\Delta x, \Delta y, \Delta z)$ is the translation of this voxel's center, $M_{\text{Pre/Post}}$ are rigid transformations that could be applied pre- and post-deformation (for DIR between 4DCT phases, these are both the $4 \times 4$ identity matrix), $(x_f, y_f, z_f)$ is the mapped location on the final image $I_f$, and all locations are specified in millimeters relative to the scanner's frame of reference. In order to compress the file size, the individual voxel locations on $I_0$ are not stored in a DICOM file, but instead must be computed from the stored minimal information as

$$\begin{bmatrix} x_0 \\ y_0 \\ z_0 \\ 1 \end{bmatrix} = \begin{bmatrix} X_{\text{Row}}\delta_X & X_{\text{Col.}}\delta_Y & X_{\text{Dep.}}\delta_Z & x_{\text{Start}} \\ Y_{\text{Row}}\delta_X & Y_{\text{Col.}}\delta_Y & Y_{\text{Dep.}}\delta_Z & y_{\text{Start}} \\ Z_{\text{Row}}\delta_X & Z_{\text{Col.}}\delta_Y & Z_{\text{Dep.}}\delta_Z & z_{\text{Start}} \\ 0 & 0 & 0 & 1 \end{bmatrix} \begin{bmatrix} i \\ j \\ k \\ 1 \end{bmatrix}, \quad (8)$$

where $(x_{\text{Start}}, y_{\text{Start}}, z_{\text{Start}})$ is the location of the starting (upper-left corner) voxel of the deformation grid, $(\delta_X, \delta_Y, \delta_Z)$ are deformation-grid resolutions, $\boldsymbol{U} = (X_{\text{Row}}, Y_{\text{Row}}, Z_{\text{Row}})$ are the row direction cosines, $\boldsymbol{V} = (X_{\text{Col.}}, Y_{\text{Col.}}, Z_{\text{Col.}})$ are the column direction cosines, $\boldsymbol{W} = (X_{\text{Dep.}}, Y_{\text{Dep.}}, Z_{\text{Dep.}})$ are the depth direction cosines such that $\boldsymbol{W} = \boldsymbol{U} \times \boldsymbol{V}$, and $(i, j, k)$ are the deformation-grid voxel indices for the given voxel. The DICOM standard also then supplies the total number of voxels in each direction $(N_X, N_Y, N_Z)$ such that $0 \leq i < N_X$ and similarly for the other dimensions. The DICOM standard additionally states that the translation for points between DVF-grid voxel centers is to be found by interpolating the surrounding grid points, *i.e.*, $T^{0 \to f}(\boldsymbol{r})$ from Eq. (1) is the trilinearly interpolated position on $I_f$ for the point $\boldsymbol{r}$ on $I_0$. Using a DVF then entails two aspects (1) calculating the location and translation at each DVF-grid location and (2) interpolating these positions to the voxel grid of interest (*e.g.*, CT or dose grids).